\pdfoutput=1

\documentclass[10pt]{article}

\usepackage{amsmath}
\usepackage{amssymb}
\usepackage{graphicx}
\usepackage{cite}
\usepackage{color} 
\usepackage{algorithm2e}

\usepackage[labelfont=bf,labelsep=period,justification=raggedright]{caption}

\makeatletter
\renewcommand{\@biblabel}[1]{\quad#1.}
\makeatother

\date{}

\newcommand{\pval}{$p$-value~}

\newcommand{\pvals}{$p$-values~}
\newcommand{\qvals}{$q$-values~}
\renewcommand{\Pr}{\mathsf{P}}
\newcommand{\bfgamma}{{\text{BF}_\gamma}}
\newcommand{\bma}{{\text{BF}_\text{BMA}}}
\newcommand{\bmalite}{{\text{BF}_\text{BMAlite}}}
\newcommand{\bmahm}{{\text{BF}^\text{HM}_\text{BMA}}}

\newcommand{\bv}{\mbox{\boldmath$b$}}

\newcommand{\ev}{\mbox{\boldmath$e$}}

\newcommand{\Cv}{\mbox{\boldmath$C$}}

\newcommand{\Gv}{\mbox{\boldmath$G$}}

\newcommand{\gv}{\mbox{\boldmath$g$}}

\newcommand{\Sv}{\mbox{\boldmath$\tau$}}

\newcommand{\lav}{\mbox{\boldmath$\lambda$}}

\newcommand{\yv}{\mbox{\boldmath$y$}}
\newcommand{\Yv}{\mbox{\boldmath$Y$}}

\newcommand{\zv}{\mbox{\boldmath$z$}}
\newcommand{\lv}{{\bf 1}}
\newcommand{\BF}{{\rm BF}}
\newcommand{\sv}{\mbox{\boldmath$s$}}
\newcommand{\cv}{\mbox{\boldmath$c$}}
\newcommand{\etv}{\mbox{\boldmath$\eta$}}
\newcommand{\wv}{\mbox{\boldmath$w$}}

\newcommand{\Wv}{\mbox{\boldmath$W$}}

\begin{document}

\begin{flushleft}
{\Large
\textbf{A statistical framework for joint eQTL analysis in multiple tissues}
}
\\
Timoth\'{e}e Flutre$^{1,2,+}$, 
Xiaoquan Wen$^{3,+}$, 
Jonathan Pritchard$^{1,4}$
Matthew Stephens$^{1,5,\ast}$
\\
\bf{1} Department of Human Genetics, University of Chicago, Chicago, IL, USA
\\
\bf{2} Department of Plant Genetics, Institut National de la Recherche Agronomique, France
\\
\bf{3} Department of Biostatistics, University of Michigan, Ann Harbor, MI, USA
\\
\bf{4} Howard Hughes Medical Institute, Chevy Chase, MD, USA
\\
\bf{5} Department of Statistics, University of Chicago, Chicago, IL, USA
\\
\bf{+} These authors contributed equally to this work.
\\
$\ast$ E-mail: mstephens@uchicago.edu
\end{flushleft}

\section*{Abstract}
Mapping expression Quantitative Trait Loci (eQTLs) 
represents a powerful and widely-adopted approach to identifying putative regulatory variants and linking them to specific genes.
Up to now eQTL studies have been conducted in a relatively narrow range of tissues or cell types.
However, understanding the biology of organismal phenotypes will involve 
understanding regulation in multiple tissues, and ongoing studies are collecting eQTL data in dozens of cell types.
Here we present a statistical framework for powerfully detecting eQTLs in multiple tissues or cell types (or,
more generally, multiple subgroups). The framework explicitly models the potential for each eQTL to be
active in some tissues and inactive in others. By modeling the sharing of active eQTLs among tissues
this framework increases power to 
detect eQTLs that are present in more than one tissue compared with ``tissue-by-tissue" analyses that examine
each tissue separately. Conversely, by modeling the inactivity of eQTLs in some tissues, the framework allows 
the proportion of eQTLs shared across different tissues 
to be formally estimated as parameters of a model, addressing the difficulties of accounting for 
incomplete power when comparing overlaps
of eQTLs identified by tissue-by-tissue analyses. Applying our framework to
re-analyze data from transformed B cells, T cells and fibroblasts we find that it substantially increases power
compared with tissue-by-tissue analysis,
identifying 63\% more genes with eQTLs (at FDR=0.05). Further the results
suggest that, in contrast to previous analyses of the same data, 
the majority of eQTLs detectable in these data are shared among all three tissues.

\section*{Author Summary}
Genetic variants that are associated with gene expression are known as expression Quantitative Trait Loci,
or eQTLs. Many studies have been conducted to identify eQTLs, and they have proven an effective tool for identifying putative regulatory variants and linking them to specific genes. Up to now most studies have been conducted
in a single tissue or cell-type, but moving forward this is changing, and ongoing studies are collecting
data aimed at mapping eQTLs in dozens of tissues. Current statistical methods are not able to
fully exploit the richness of these kinds of data, taking account of both the sharing and differences in eQTLs among tissues. 
In this paper we develop a statistical framework to address this problem, to improve power to detect
eQTLs when they are shared among multiple tissues, and to allow for differences among tissues to be estimated.
Applying these methods to data from three tissues suggests that sharing of eQTLs among tissues may 
be substantially more common than it appeared in previous analyses of the same data.

\section*{Introduction}
Regulatory variation plays an essential role in the genetics of disease and other phenotypes as well as in evolutionary change \cite{Frazer2009Human,Montgomery2011From,Wray2007Evolutionary}.
However, in sharp contrast to nonsynonymous variants in the human genome, which can now be identified with great accuracy,
it remains extremely difficult to know which variants in the genome may impact gene regulation in any given tissue or cell type.
[Henceforth we use ``tissue" for brevity, but everything applies equally to cell types.]
Expression QTL mapping (e.g.~\cite{Cheung2003Natural,Stranger2007Population,Gilad2008Revealing} represents a powerful approach for bridging this gap, by allowing regulatory variants to be identified, and linked to specific genes.
Indeed, numerous studies (e.g., \cite{Nica2010Candidate,Nicolae2010Traitassociated}) have shown highly significant overlaps between eQTLs and SNPs associated with organismal-level phenotypes in genome-wide association studies (GWAS), suggesting that a large fraction of GWAS associations may be due to variants that affect gene expression.

Ultimately, understanding the biology of organismal phenotypes, such as diseases, is likely to require understanding regulatory variation in many different tissues (\cite{Cookson2009Mapping,Greenawalt2011Survey}).
For example, if regulatory variants differ across tissues, then, in understanding GWAS hits, and using them to understand the biology of disease, we would like to know which variants are affecting which tissues.
At a more fundamental level, identifying differential genetic regulation in different tissues could yield insights into the basic biological processes that influence tissue differentiation.
To date, eQTL studies have been performed in a relatively narrow range of tissue types.
However, this is changing quickly: for example, the NIH ``Genotype-Tissue Expression'' (GTEx) project aims to collect expression and genotype data in 30 tissues across 900 individuals. 
Motivated by this, here we describe and illustrate a statistical framework for mapping eQTLs in expression data on multiple tissues.

While statistical methods for identifying eQTLs in a single tissue or cell type  are now relatively mature (e.g.~\cite{Stegle2012Using}) current analytic tools are limited in their ability to fully exploit the richness of data across multiple tissues.
In particular, available methods fall short in their ability to {\it jointly analyze data on all tissues} to maximize power, while simultaneously {\it allowing for differences among eQTLs} present in each tissue.
Indeed relatively few papers have considered the problem.
The simplest approach (e.g. \cite{Dimas2009Common,Nica2011Architecture}) is to analyze data on each tissue separately (``tissue-by-tissue" analysis), and then to examine overlap of results among tissues.
However, this fails to leverage commonalities among tissues to improve power to detect shared eQTLs.
Furthermore, although examining overlap of eQTLs among tissues may appear a natural approach to examining heterogeneity, in practice interpretation of results is complicated by the difficulty of accounting for incomplete power.
Both \cite{Ding2010Gene} and \cite{Nica2011Architecture} provide approaches to address this, but only for pairwise comparisons of tissues.  

Compared with tissue-by-tissue analysis, joint analysis of multiple tissues has the potential to increase power to identify eQTLs that have similar effects across tissues.
Both \cite{Gerrits2009Expression} and \cite{Fu2011Unraveling} conduct such joint analyses -- the first using ANOVA, and the second using a weighted $Z$-score meta-analysis -- and \cite{Fu2011Unraveling} confirm that their joint analysis has greater power than tissue-by-tissue analysis.
The ANOVA and $Z$-score methods each have different advantages.
The ANOVA framework has the advantage that, by including interaction terms, it can be used to investigate heterogeneity in eQTL effects among tissues.
Gerrits \textit{et al.}\ (\cite{Gerrits2009Expression}) use this to identify eQTLs that show significant heterogeneity, and then classify these eQTLs, post-hoc, into different types based on estimated effect sizes. 
The weighted $Z$-score method has the advantage that, unlike ANOVA, it allows for different variances of expression levels in different tissues (which are likely to occur in practice).
However, it does not so easily allow for investigation of heterogeneity; Fu \textit{et al.}\ (\cite{Fu2011Unraveling}) assess heterogeneity for pairs of tissues by using a resampling-based procedure to assess the significance of observed differences in $Z$ scores.

Here we introduce a statistical framework for the joint analysis of eQTLs among multiple tissue types, that combines advantages of some of the methods above, as well as introducing some new ones.
In brief, our framework integrates recently-developed GWAS meta-analysis methods that allow for heterogeneity of effects among groups \cite{Lebrec2010Dealing,Han2011RandomEffects,Wen2011Bayesian,Han2012Interpreting}, into a hierarchical model (e.g.~\cite{Veyrieras2008Highresolution,Maranville2011Interactions}) that combines information across genes to estimate the relative frequency of patterns of eQTL sharing among tissues.
Like ANOVA, our approach allows investigation of heterogeneity among several tissues, not just pairs of tissues.
However, in contrast to ANOVA, our framework allows for different variances in different tissues.
Moreover, unlike any of the methods described above, our framework explicitly models the fact that some tissues may share eQTLs more than others, and estimates these patterns of sharing from the data.
Our methods also allow for intra-individual correlations when samples are obtained from a common set of individuals.
While we focus here on comparing and combining information across 
different tissue types, our framework could be applied equally to 
comparing and combining across other units, e.g. different experimental platforms, multiple datasets on the same tissue types, or data
on individuals from different populations.

The remainder of the paper is as follows.
After providing a brief overview of our framework, we use simulations to illustrate its power compared to other methods, and then apply it to map eQTLs, and assess heterogeneity among tissues, using data from Fibroblasts, LCLs and T-cells (\cite{Dimas2009Common}).
Consistent with results from \cite{Fu2011Unraveling}, we show that our joint analysis framework provides a large gain in power compared with a tissue-by-tissue analysis.
Furthermore,  compared with previous analyses of these data, we find a much higher rate of tissue-consistent eQTLs.

\section*{Results}
\subsection*{Methods Overview}

Consider mapping eQTLs in $S$ tissues. In our applications here
the expression data are from micro-arrays, and so we assume a normal
model for the expression levels, suitably-transformed.  (These methods can also be applied to RNA-seq data after suitable transformation; see Discussion). That is, 
in each tissue, $s=1,\dots,S$, we model the potential association between a candidate SNP and a target gene by a linear regression:
\begin{equation}
  \label{simple.reg.model}
  y_{si} = \mu_s + \beta_s g_i + \epsilon_{si} \; \text{ with }\epsilon_{si} \sim \mathcal{N}(0,\sigma_s^2),
\end{equation}
where $y_{si}$ denotes the observed expression level of the target gene in tissue $s$ for the $i^{th}$ individual, $\mu_s$ the mean expression level of this gene in tissue $s$, $\beta_s$ the effect of a candidate SNP on this gene expression in tissue $s$, $g_i$ the genotype of the $i^{th}$ individual at the SNP (coded as 0,1 or 2 copies of a reference allele) and $\epsilon_{si}$ the residual error for tissue $s$ and individual $i$.
Note that the subscript $s$ on residual variance $\sigma^2_s$ indicates that we allow the residual variance to be different in each tissue.
In addition, when tissues are sampled from the same set of individuals,
we allow that the residual errors $\epsilon_{1i},\dots,\epsilon_{Si}$ may be correlated (with the correlation matrix to be estimated from the data).

The primary questions of interest are whether the SNP is an eQTL in any tissue, and, if so, in which tissues.
To address these questions we use the idea of a ``configuration" from \cite{Wen2011Bayesian,Han2012Interpreting}.
A configuration is a binary vector $\gamma=(\gamma_1,\dots,\gamma_S)$ where $\gamma_s \in \{0,1\}$ indicates whether the SNP is an eQTL in tissue $s$.
If $\gamma_s=1$ then we say the eQTL is ``active" in tissue $s$.
The ``global null hypothesis", $H_0$, that the SNP is not an eQTL in any tissue, is therefore $\gamma=(0,\dots,0)$.
Every other possible value of $\gamma$ can be thought of as representing a particular alternative hypothesis.
For example, $\gamma=(1,\dots,1)$ represents the alternative hypothesis that the SNP is an eQTL in all $S$ tissues, and $\gamma=(1,0,\dots,0)$ represents the alternative hypothesis that the SNP is an eQTL in just the first tissue.

Our aim is to perform inference for $\gamma$.
A natural approach is to specify a probability model, $\Pr(\text{data } | \, \gamma)$, being the probability of obtaining the observed data if the true configuration were $\gamma$, and then perform likelihood-based inference for $\gamma$. 
The support in the data for each possible value of $\gamma$, relative to the null $H_0$, is quantified by the likelihood ratio, or \emph{Bayes Factor} (BF, \cite{Kass1995Bayes}):
\begin{equation} \label{eqn:bfgamma}
\bfgamma = \frac{\Pr(\text{data } | \text{ true configuration is $\gamma$})}{\Pr(\text{data } | \, H_0)}.
\end{equation}
Specifying these likelihoods requires assumptions about $\Pr(\beta | \gamma)$, the distribution of the effect sizes $\beta$ for each possible configuration $\gamma$ (as well as less crucial assumptions about nuisance parameters such as $\mu$ and $\sigma_s$). 
Of course, if $\gamma_s=0$ then $\beta_s=0$ by definition, but for the tissues where $\gamma_s=1$ various assumptions are possible -- for example, one could assume that the effect $\beta_s$ is the same in all these tissues, or allow it to vary among tissues.
Here we use a flexible family of distributions, $\Pr(\beta | \gamma, \theta)$ (see Methods), where the hyper-parameters $\theta$ can be varied to control both the typical effect size, and the heterogeneity of effects across tissues (see below). 

The value of $\bfgamma$ measures the support in the data for one specific alternative configuration $\gamma$, compared against the null hypothesis $H_0$. 
To account for the fact that there are many possible alternatives, the {\it overall} strength of evidence against $H_0$ at the candidate SNP
is obtained by ``Bayesian Model Averaging" (BMA), which involves averaging $\bfgamma$ over the possible alternative configurations $\gamma$, weighting each by its prior probability, $\eta_\gamma$: 
\begin{equation} \label{eqn:bma}
\bma = \frac{\Pr(\text{data } | \text{ $H_0$ false})}{\Pr(\text{data } | \text{ $H_0$ true})} = \sum_\gamma \, \eta_\gamma \, \bfgamma.
\end{equation}
Further, under an assumption of at most one eQTL per gene, the overall evidence against $H_0$ for the entire gene (i.e. that the gene contains no eQTL in any tissue) 
is given by averaging $\bma$ across all candidate SNPs \cite{Servin2007Imputationbased}.
In either case, at either the SNP or gene level, large values of $\bma$ constitute strong evidence against $H_0$.
$\bma$ has a direct Bayesian interpretation as the strength of the evidence against $H_0$, but here we also use it as a frequentist test statistic (\cite{Good1992BayesNonBayes,Servin2007Imputationbased}), assessing significance by permutation or simulation.
The latter has the advantage that \pvals and \qvals obtained in this way are ``valid" even if not all the prior assumptions are exactly correct.

Note that $\bma$ depends on the choice of $(\theta,\eta)$, and the power of $\bma$ as a test statistic is expected to depend on how well this choice of these hyper-parameters captures the range of alternative scenarios present in the data.
Here we make use of three different choices:
\begin{itemize}
\item A ``data-driven" choice, where the hyper-parameters are estimated from the data using a hierarchical model (HM, \cite{Gelman2006Data}) that combines
information across all genes. We use $\bmahm$ to denote this choice.
\item A ``default" choice, which chooses $\eta$ to cover a wide range of different possible alternative configurations, and $\theta$ is set to allow modest heterogeneity. We use $\bma$ to denote this choice.
\item A ``lite" choice, which puts weight only on the most extreme configurations (where the eQTL is active in only one tissue, or in all tissues), but compensates by setting $\theta$ to allow for more heterogeneity. We use $\bmalite$ to denote this choice.
\end{itemize}
Each of these choices has something to recommend it.
The first, being data driven, is the most attractive in principle, but also the most complex to implement.
The default choice is simpler to implement, and is included partly to demonstrate that one does not have to get the hyper-parameter values exactly ``right" for $\bma$ to be a powerful test statistic. 
Finally, $\bmalite$ has the advantage that it is easily applied to large numbers of tissues; neither of the other methods scales well, either computationally or statistically, with the number of tissues, because the number of terms in the sum in equation (\ref{eqn:bma}) is $2^S-1$.

When there is strong evidence against $H_0$, the Bayes Factors can also be used to assess which alternative configurations $\gamma$ are consistent with the data.
Specifically the posterior probability on each configuration is:
\begin{equation} \label{eqn:config}
\Pr(\text{true configuration is $\gamma$} \, | \text{ data, $H_0$ false}) = \frac{\eta_\gamma \text{BF}_\gamma}{\sum_\gamma \eta_\gamma \text{BF}_\gamma},
\end{equation}
and the posterior probability that the SNP is an eQTL in tissue $s$ is obtained by summing the probabilities over configurations in which $\gamma_s=1$:
\begin{equation} \label{eqn:tissueprob}
\Pr(\text{eQTL in tissue $s$} \, | \text{ data, $H_0$ false}) = \sum_{\gamma: \gamma_s=1} \Pr(\text{true configuration is $\gamma$ } | \text{ data, $H_0$ false}).
\end{equation}
The second of these is particularly helpful when the data are informative for an eQTL in tissue $s$, but ambiguous in other tissues: in such a case the probability $(\ref{eqn:tissueprob})$ will be close to 1, even though the ``true" configuration will be uncertain (so none of the probabilities (\ref{eqn:config}) will be close to 1).
Because both (\ref{eqn:config}) and (\ref{eqn:tissueprob}) are sensitive to choice of hyper-parameters, we compute them using $\bmahm$ (where the hyper-parameters are estimated from the data).

Further details of methods used are provided in the Methods section.


\subsection*{Simulations}

\subsubsection*{Power to detect eQTLs}

We begin by comparing the ability of different methods to reject the global null hypothesis $H_0$; i.e. to detect eQTLs that occur in {\it any} tissues.
We expect that a tissue-by-tissue analysis, which analyzes each tissue separately, will perform well for detecting eQTLs that are present in a single tissue.
Conversely, we expect joint analysis of all tissues to perform well for detecting eQTLs that are present in all tissues.
Our Bayesian model averaging (BMA) approach attempts, by averaging over different possible eQTL configurations, to combine the advantages of both types of analysis, and  thus aims to perform well across all scenarios. 

To assess this we performed a series of simulations, with five tissues measured in 100 individuals (and no intra-individual correlations).
Each simulation consisted of 2,000 gene-SNP pairs (i.e.~one candidate SNP per gene), half of which were ``null" (i.e.~the SNP was not an eQTL in any tissue), and the other half following an alternative hypothesis where the SNP was an eQTL in exactly $k$ tissues, with $k$ varying from 1 to 5.  
Thus, for example, the simulations with $k=1$ assess power to detect eQTLs that are active in just one tissue, whereas the simulations with $k=5$ assess power to detect eQTLs that are active in all five tissues. 
When simulating eQTLs that are active in multiple tissues we assumed their effects to be similar, but not identical, across tissues (see Methods).
We applied four analysis methods to these data:
1) $\bma$, being our Bayesian Model Averaging approach with default weights described above; 2) $\bmalite$, being the computationally-scalable version of BMA described above;
3) ANOVA/linear regression (ANOVA/LR) (c.f.~\cite{Gerrits2009Expression} and see Methods), which jointly analyzes all tissues in a regression model, and compares the general alternative model (which allows a different genetic effect in each tissue) with the null model (no effect in all tissues);
and 4) a ``tissue-by-tissue" analysis (c.f.~\cite{Dimas2009Common}), where we use linear regression to test for an eQTL separately in each tissue, and take the minimum \pval across tissues as a test statistic.
For simplicity we defer consideration of the more sophisticated of our approaches, $\bmahm$, to slightly more complex simulations described later.
Each of these methods produces a test statistic for each SNP-gene pair, testing the global null hypothesis $H_0$.
For each test statistic, we found the threshold that yielded a False Discovery Rate of 0.05 (based on the known null/alternative status of each SNP-gene pair), and assessed the effectiveness of each method by the number of discoveries it made at that FDR.

\begin{figure}[!ht]
\begin{center}
\includegraphics[width=\textwidth]{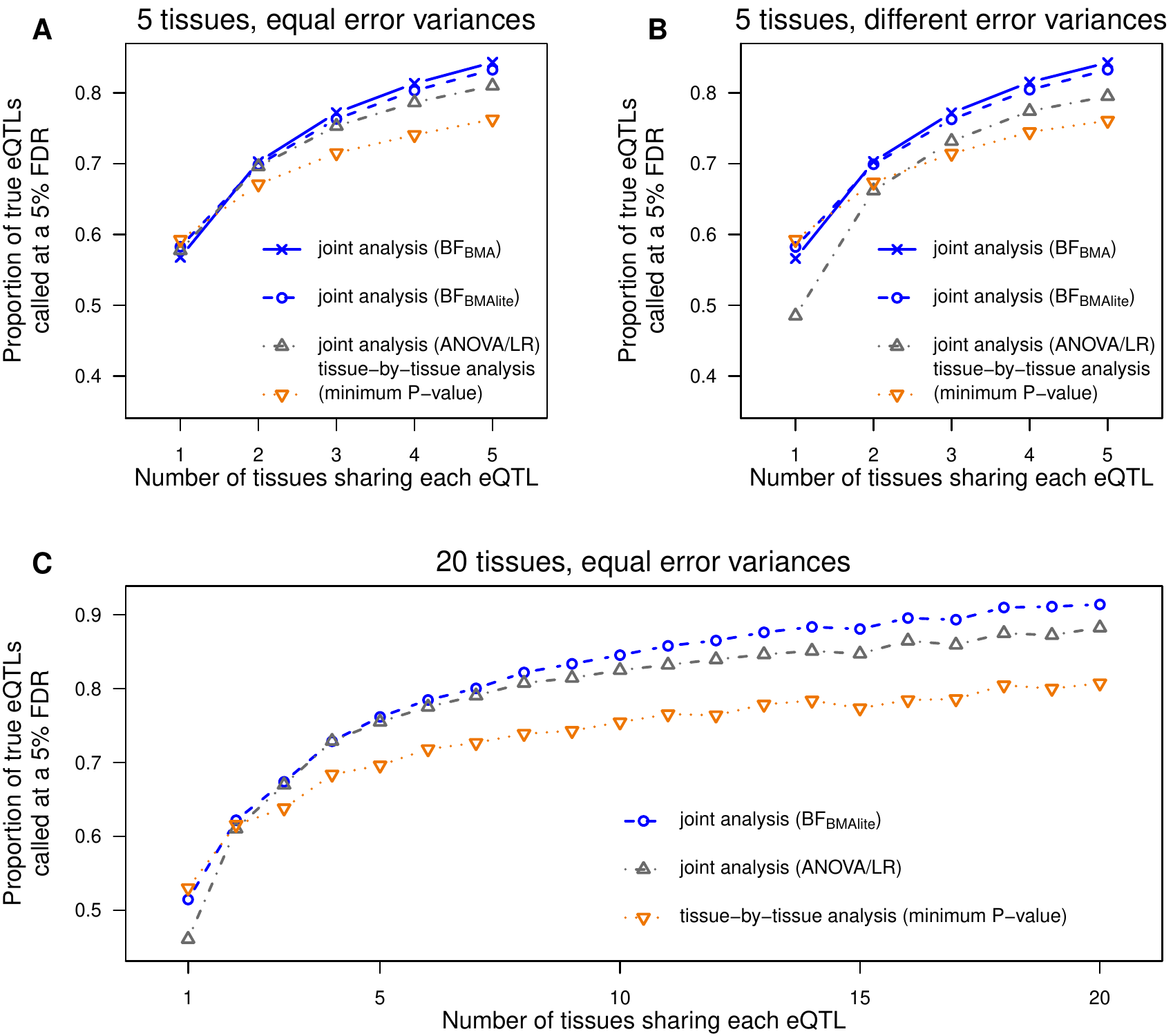}
\end{center}
\caption{
{\bf The $\bma$ joint analysis has more power across a range of alternatives.} A. Five tissues are simulated, each with the error variance equal to 1. B. Five tissues are simulated, with error variances being 1, 1.5 or 2. C. Twenty tissues are simulated, each with the error variance equal to 1.
}
\label{fig:fig1}
\end{figure}

The results of these comparisons are shown in Figure \ref{fig:fig1}A.
As expected, for eQTLs that occur in just one tissue, the tissue-by-tissue analysis performs best.
However, it is only slightly more effective than the joint analysis approaches in this setting.
Conversely, the joint analysis approaches outperform the tissue-by-tissue analysis for eQTLs that occur in more than one tissue, with the gains becoming larger as the number of tissues sharing the eQTL increases.
The BMA analyses generally perform similarly to one another, and outperform ANOVA/LR. This is presumably because our simulations involved eQTLs that have similar effects in each tissue, and our prior distribution $p(\beta | \gamma,\theta)$ explicitly up-weights eQTLs with this feature. 

This first set of simulations assumed error variances to be equal among tissues.
This assumption is made by ANOVA/LR, but not by the other methods, and is likely often to be violated in practice.
To assess the effects of this we repeated the simulations, but with error variances differing among tissues.
The results (Figure \ref{fig:fig1}B) confirm that, relative to other methods, ANOVA/LR performs less well when error variances vary among tissues.

To assess performance in larger numbers of tissues we repeated the simulations above, but with 20 tissues (so $k=1,\dots,20$).
For this many tissues computing $\bma$ involves averaging over all $2^{20}>10^6$ possible eQTL configurations, which is computationally inconvenient,
so we omitted $\bma$ from this comparison.
The results (Figure \ref{fig:fig1}C) show that $\bmalite$ performs similarly to the tissue-by-tissue analysis for eQTLs that occur in just one or two tissues, and outperforms it substantially for eQTLs occurring in many tissues. As expected, ANOVA/LR outperforms tissue-by-tissue analysis for eQTLs occurring in many tissues, but is noticeably less effective for eQTLs occurring in
only one tissue, and performs consistently less well than $\bmalite$. 

In summary, these simulations illustrate the benefits of Bayesian Model Averaging as a general strategy for producing powerful test statistics: by explicitly averaging over a range of alternative models, 
these test statistics are able to achieve good power to detect a wide range of different types of eQTL.

\subsubsection*{Identifying eQTLs in particular tissues: borrowing information among tissues}

Next we consider the benefits of jointly analyzing multiple tissues, even when the main goal is to identify eQTLs in a particular tissue of interest.
For intuition, suppose for a moment that every eQTL is shared among all tissues. Then, from the simulation results above, we know that a joint analysis will identify more eQTLs overall, and hence more eQTLs in the tissue of interest.
Of course, not all eQTLs are shared among all tissues, but some are, and some tissues may share eQTLs more than others.
To allow for this, our hierarchical model attempts to infer the extent of such sharing (by estimating the configuration weights $\eta_\gamma$), and exploits any sharing that does occur to increase power to detect eQTLs in each tissue.
By estimating sharing from the data, rather than assuming that all tissues share equally with one another (as do the simpler test statistics $\bma$ and $\bmalite$ used above), we expect 
$\bmahm$ to make more effective use of sharing in the data to further improve power to identify eQTLs.

To illustrate this, we simulated eQTL data for five tissues.
Some eQTLs were shared by all tissues, some were specific to each tissue, some were shared by Tissues 1 and 2 only, and some were shared by Tissues 3, 4 and 5.
To show how the benefits of sharing can change with sample size, we simulated 60 samples for Tissue 1, and 100 samples for the others.
This mimics a setting where Tissue 1 is harder to obtain than the other tissues, with Tissue 2 being the best proxy for Tissue 1. 

We applied our Bayesian methods and a tissue-by-tissue analysis to these data, and assessed their ability to identify eQTLs in each tissue.
For the tissue-by-tissue analysis the test statistic in each tissue is simply the linear regression \pval in that tissue. 
For our Bayesian methods, the test statistic in each tissue is the posterior probability of the SNP being an active eQTL in that tissue (\ref{eqn:tissueprob}).
Note that this posterior probability is computed from joint analysis of all tissues, and takes account of sharing of eQTLs among tissues.
For example, consider a SNP showing modest association with expression in Tissue 1.
If this SNP also shows strong association in the other tissues, then it will be assigned a higher probability of being an active eQTL in Tissue 1 than it would if it showed no association in the other tissues.
For each method, separately in each tissue, we identified the threshold of the test statistic value that yields a FDR of 0.05 in that tissue, based on the true active/inactive status of each SNP in that tissue (known since this is simulated data).
(A SNP that is an eQTL in some tissues but not others counts as a ``false discovery" if it is called as an eQTL in a tissue where it is inactive.)
For the Bayesian methods we obtained results both using ``default" weights on configurations ($\bma$), and using weights estimated from the data by the hierarchical model ($\bmahm$).
The latter is expected to be more effective as it should learn, for example, that Tissue 1 shares more eQTLs with Tissue 2 than with other tissues.

\begin{figure}[!ht]
\begin{center}
\includegraphics[width=\textwidth]{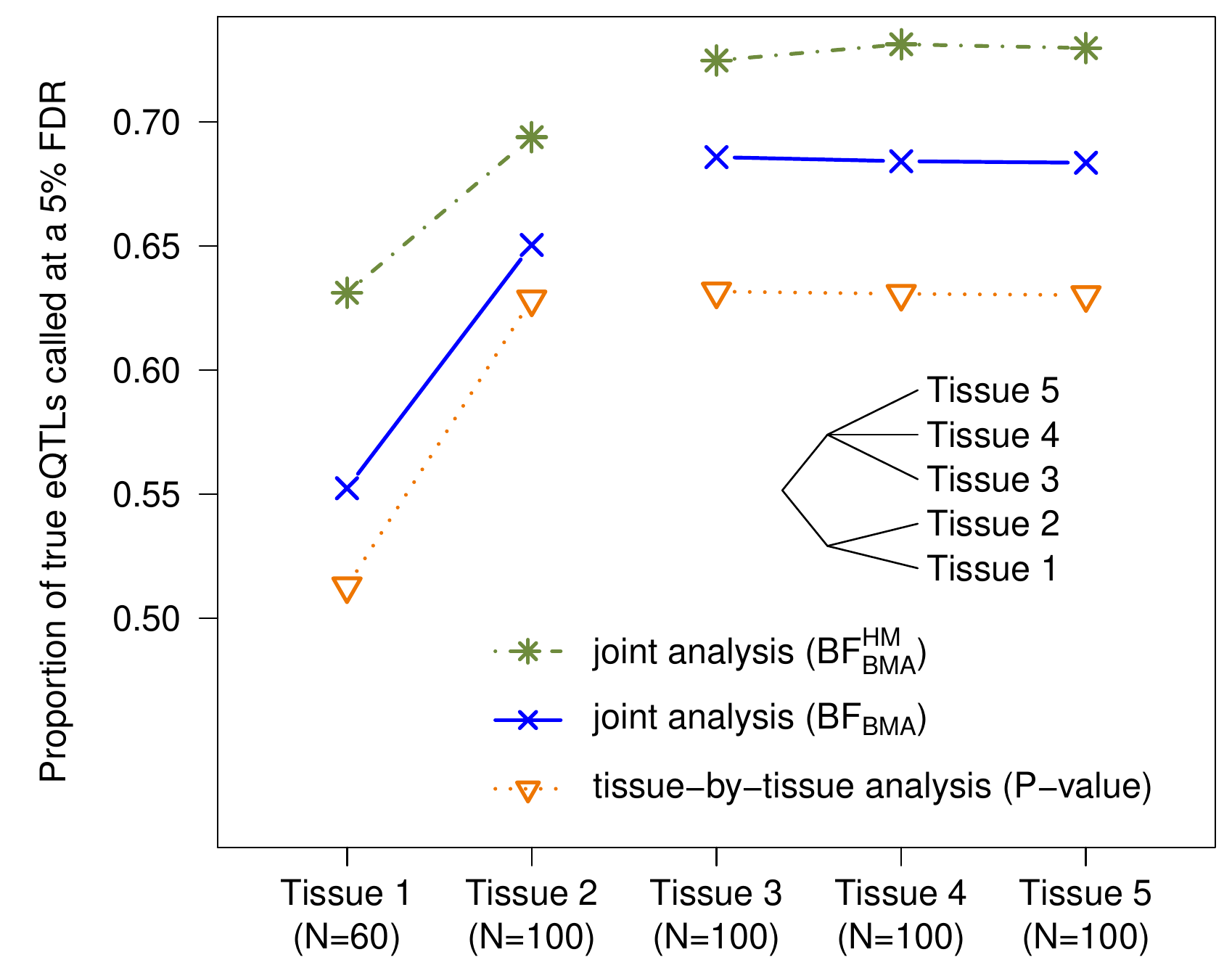}
\end{center}
\caption{
{\bf The $\bmahm$ joint analysis efficiently borrows information across genes.} Five tissues are simulated. Some eQTLs were shared by all tissues,
some were specific to each tissue, and, as depicted by the cladogram, some were shared by Tissues 1 and 2 only, while others were shared by Tissues 3, 4 and 5. Each tissue has 100 samples, except tissue 1 which has only 60.
}
\label{fig:fig2}
\end{figure}

The results (Figure \ref{fig:fig2}) show that, for all tissues, our joint analyses outperform the tissue-by-tissue analysis.
Further, $\bmahm$ outperforms $\bma$, demonstrating the benefits of learning patterns of sharing from the data.
Finally, the gain of $\bmahm$ is greater for Tissue 1 than for Tissue 2, illustrating that benefits of sharing information are greater for tissues with small sample sizes.

Furthermore, using the hierarchical model which pools all genes together, we can estimate the configuration proportions, i.e.~$\eta_\gamma$.
In the setting described above, we simulated one thousand eQTLs in each of 8 different configurations, as well as one thousand genes with no eQTLs.
Averaged over replicates, the proportions are estimated to be in $[0.124-0.127]$ for each of the 8 active configurations (negligible differences between replicates).
These estimates are fairly accurate knowing that the true proportion is $1/8 = 0.125$ for each configuration.


\subsection*{Analysis of eQTL data in three cell types from Dimas \textit{et al.}}

We now analyze data from \cite{Dimas2009Common}, consisting of gene expression levels measured in fibroblasts, LCLs and T-cells from 75 unrelated individuals genotyped at approximately 400,000 SNPs. 
The data were pre-processed similarly to the original publication, as described in the Methods section.
Throughout we focus on testing SNPs that lie within 1Mb of the transcription start site of each gene (the ``\textit{cis} candidate region"), and on a subset of 5012 genes robustly expressed in all
three cell-types.

\subsubsection*{Gain in power from joint analysis}

First we assess the gain in power from mapping eQTLs in all three cell types jointly, using $\bma$, compared with a ``tissue-by-tissue" analysis similar to that in \cite{Dimas2009Common}.
For each method we compute a test statistic for each gene, combining information across SNPs, to assess the overall support for any eQTL in that gene in any tissue. 
For our Bayesian approach the test statistic is the average value of $\bma$ over all SNPs in the cis candidate region; 
for the tissue-by-tissue analysis the test statistic is the minimum \pval from linear regressions performed separately in each tissue for each SNP in the cis candidate region.
We translate each of these test statistics into a \pval for each gene by comparing observed values with simulated values obtained under $H_0$ (by permutation of individual labels).
Finally we computed, for each method, the number of genes identified as having an eQTL at an FDR of 0.05 (using the \verb+qvalue+ package \cite{Storey2003Statistical}).

\begin{figure}[!ht]
\begin{center}
\includegraphics[width=\textwidth]{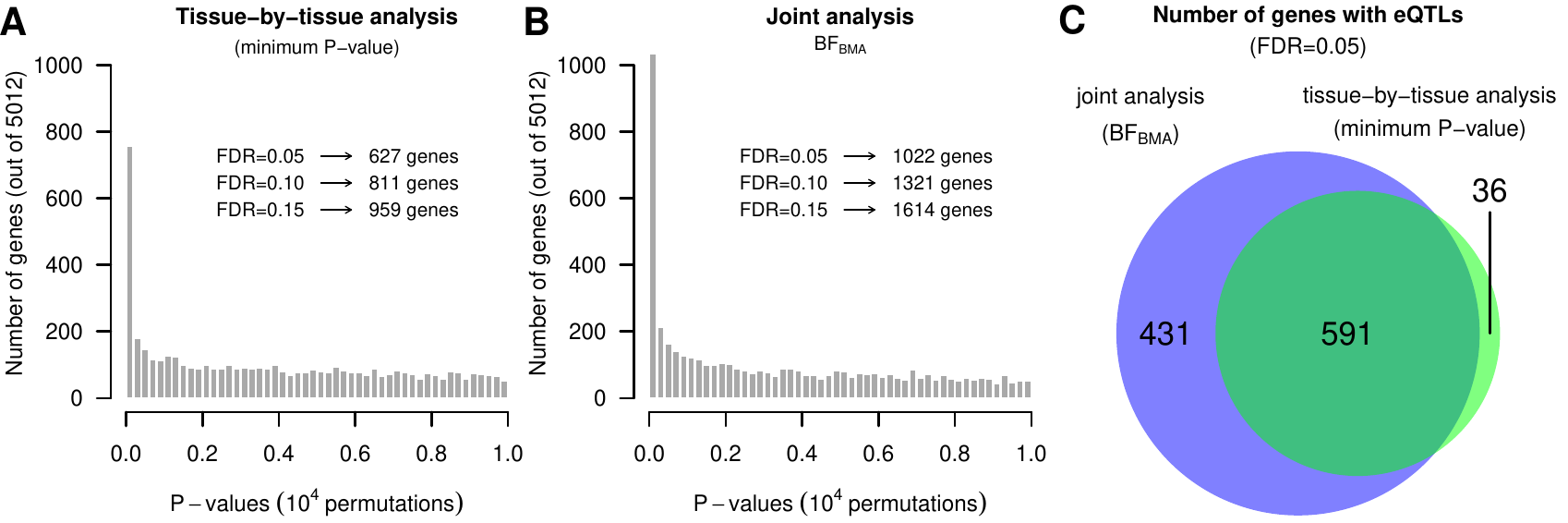}
\end{center}
\caption{
{\bf The $\bma$ joint analysis is more powerful on the data set from Dimas \textit{et al}.} A and B. Histograms of gene \pvals obtained by the tissue-by-tissue analysis and the joint analysis. C. Numbers of eQTLs called by both methods or either one of them.
}
\label{fig:fig3}
\end{figure}

Joint mapping, via $\bma$, substantially increased power to identify eQTLs compared with tissue-by-tissue analysis. For example,
$\bma$ identified 1022 eQTLs at FDR=0.05, which is $63\%$ more than the 627 eQTLs identified by the tissue-by-tissue analysis at the same FDR (Figure \ref{fig:fig3} A and B).
Further, the vast majority of eQTLs identified by the tissue-by-tissue analysis ($94\%$) are also detected by $\bma$ (Figure \ref{fig:fig3}C).

\begin{figure}[!ht]
\begin{center}
\includegraphics[width=\textwidth]{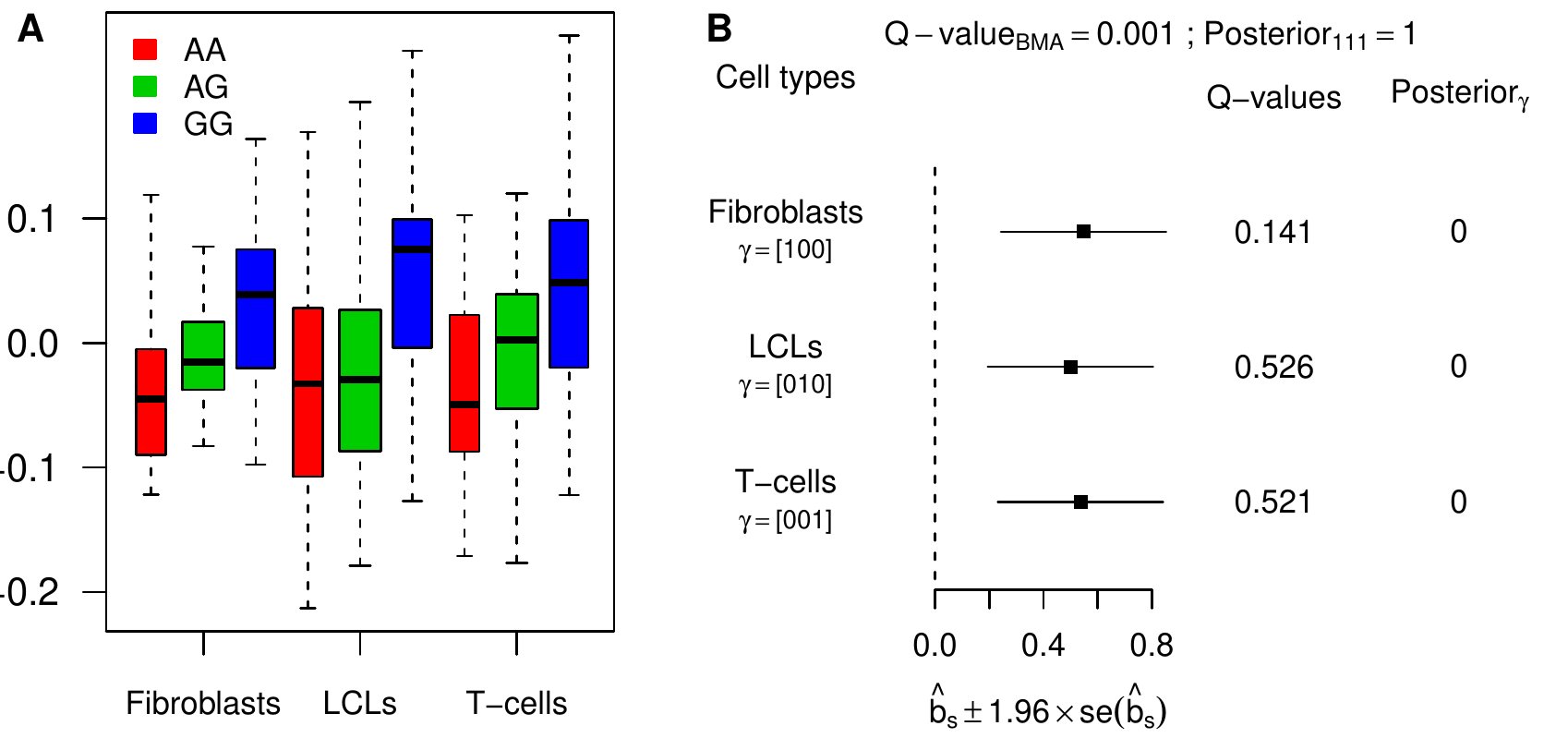}
\end{center}
\caption{
{\bf Example of an eQTL with weak, yet consistent effects.}
A. Boxplots of the PC-corrected expression levels from gene \textit{ASCC1} (Ensembl id ENSG00000138303) in all three cell types, color-coded by genotype class at SNP rs1678614.
B. Forest plot of estimated standardized effect sizes of this eQTL. Note that none of the \qvals from the tissue-by-tissue analysis are significant at FDR=0.05.
}
\label{fig:fig4}
\end{figure}

In many cases, the eQTLs detected by $\bma$ but not by the tissue-by-tissue analysis have modest effects that are consistent across tissues.
Because their effects are modest in each tissue, they fail to reach the threshold for statistical significance in any single tissue, and so the tissue-by-tissue analysis misses them.
But because their effects are consistent across tissues, the joint analysis is able to detect them.
Figure \ref{fig:fig4} shows an example illustrating this (gene {\it ASCC1}, Ensembl id ENSG00000138303, with SNP rs1678614).
The PC-corrected phenotypes already indicate that this gene-SNP pair looks like a consistent eQTL (Figure \ref{fig:fig4}A), and
its effect size estimates are highly concordant across tissues (Figure \ref{fig:fig4}B).
However, as indicated by the \qvals on the forest plot, this eQTL is not called by the tissue-by-tissue analysis in any tissue (all the \qvals exceed $.14$).
In contrast, the joint analysis pools information across tissues to conclude that there is strong evidence for an eQTL ($q=0.001$),
and that it is likely an eQTL in all three tissues (probability 1 assigned to the consistent configuration $\gamma=[111]$, conditional on it being an eQTL).

\subsubsection*{Many eQTLs are consistent among tissues}

The original analyses of these data concluded that  
69\% to 80\% of eQTLs operate in a cell-type specific manner (\cite{Dimas2009Common}). These results were obtained by mapping eQTLs separately in each tissue,
and then examining which of the eQTLs identified in each tissue also showed some signal (e.g.~at a relaxed significance threshold of $p=0.05$), 
in another tissue. However, as noted by \cite{Ding2010Gene}, due to incomplete power, eQTLs that are actually shared between tissues
may appear ``tissue-specific" in this type of analysis. Our hierarchical model has the potential to help overcome this difficulty
by estimating the proportion of eQTLs that follow each configuration type as a parameter of the model, combining information across
all genes, and without setting specific significance thresholds (thus sidestepping the problems of incomplete power).

Applying the hierarchical model to these data produced an estimate of just $8\%$ of eQTLs being specific to a single tissue, 
with an estimated $88\%$ 
of eQTLs being common to all three tissues (95\% CI = 84\%-93\%; Table \ref{tab:tab1}). 
Among eQTLs shared between just two tissues, many more are shared between LCLs and T-cells, than between these cell types and fibroblasts.
This is consistent with results from \cite{Dimas2009Common}, and not unexpected since LCLs and T-cells are 
more similar to one another than to fibroblasts.  

\begin{table}[!ht]
\caption{
\bf{Inference of the proportion of tissue specificity}}
\centering
\begin{tabular}{|c|c|c|}
Configuration & Hierarchical model & Tissue-by-tissue \\
F-L-T & 0.882 [0.840, 0.925] & 0.187 \\
L-T & 0.051 [0.025, 0.085] & 0.080 \\
F-L & 0.005 [0.000, 0.018] & 0.050 \\
F-T & 0.002 [0.000, 0.011] & 0.047 \\
F & 0.033 [0.014, 0.065] & 0.246 \\
L & 0.015 [0.000, 0.039] & 0.165 \\
T & 0.011 [0.000, 0.033] & 0.224 \\
\end{tabular}
\begin{flushleft}The results for the hierarchical model were obtained with the multivariate Bayes Factors allowing correlated residuals and the EM algorithm. The results for the tissue-by-tissue analysis were obtained by calling eQTLs at an FDR of 0.05 after performing permutations in each tissue separately, and calculating the overlaps among tissues.
\end{flushleft}
\label{tab:tab1}
\end{table}

We obtained qualitatively similar patterns when we varied some of the assumptions in the hierarchical model - 
specifically, whether or not we allow for intra-individual correlations, whether or not we assume at most one eQTL per gene,
whether or not we remove PCs to account for confounders, and whether or not we analyze all genes or only those genes robustly expressed in all tissues (Supplementary text S1). 
Nonetheless, we caution against putting too much weight on any particular number to quantify tissue specificity, not least because
the definition of a tissue-specific eQTL is somewhat delicate: for example, it is unclear how to classify a
SNP that is very strong eQTL in one tissue, and much weaker in the others. Further, our estimates necessarily
reflect patterns of sharing only for moderately strong eQTLs, strong enough to be detected in the modest
sample sizes available here: patterns of sharing could be different among weaker eQTLs. Nonetheless,
these results do suggest that there is substantial sharing of eQTLs among these three tissue
types, considerably higher than the original analysis suggested.

\begin{figure}[!ht]
\begin{center}
\includegraphics[width=\textwidth]{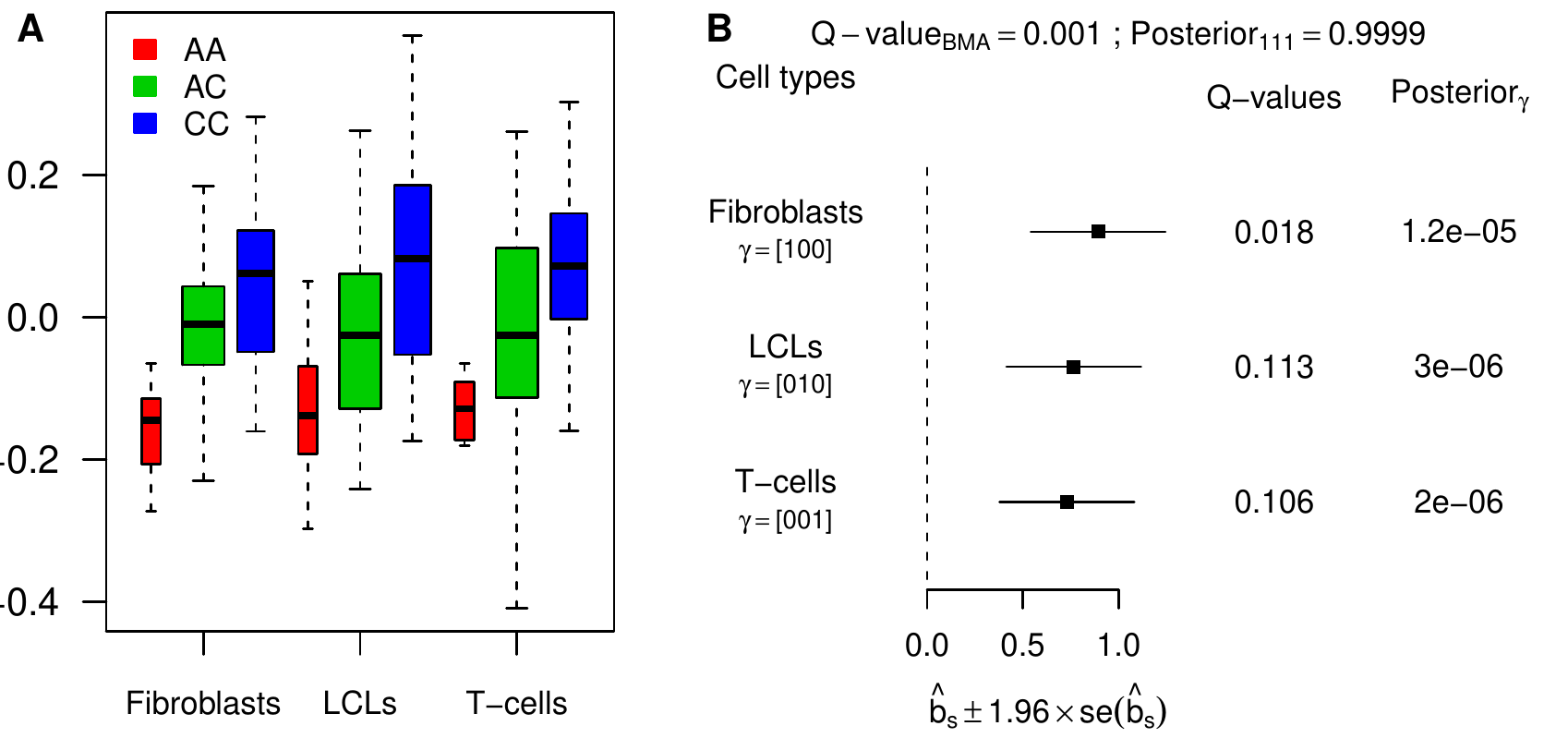}
\end{center}
\caption{
{\bf Example of an eQTL wrongly called as tissue-specific by the tissue-by-tissue analysis.}
A. Boxplots of the PC-corrected expression levels from gene \textit{CHPT1} (Ensembl id ENSG00000111666) in all three cell types, color-coded by genotype class at SNP rs10860794.
B. Forest plot of estimated standardized effect sizes of this eQTL. Note that, from the \qvals of the tissue-by-tissue analysis, the eQTL is significant at FDR=0.05 only in fibroblasts.
}
\label{fig:fig5}
\end{figure}

To illustrate the potential pitfalls of investigating heterogeneity in a tissue-by-tissue analyses,
we also ran a tissue-by-tissue analysis on these data. Specifically, we called eQTLs separately in each tissue (at an FDR of 0.05), and then 
examined the overlap in the genes identified in each tissue. Using this procedure, in strong contrast
with results from the joint analysis, $65\%$ of eQTLs are called in only one tissue, with fewer than
15\% called in all three tissues (Table \ref{tab:tab1}).
Qualitatively similar results are obtained for different FDR thresholds.
However, these results cannot be taken as reliable indications of tissue specificity,
because the procedure fails to take account of incomplete power to detect eQTLs at any given threshold,
 and therefore tends to over-estimate tissue specificity.
Figure \ref{fig:fig5} shows an eQTL that illustrates this behavior (gene {\it CHPT1}, Ensembl id ENSG00000111666, with SNP rs10860794).
Visual examination of the expression levels in each genotype class (Figure \ref{fig:fig5}A), suggest that this SNP is an
eQTL in all three tissues, with similar effects in each tissue (Figure \ref{fig:fig5}B). This is supported by the joint analysis,
which shows strong evidence for an eQTL $q=0.001$, and assigns probability effectively 1 to the consistent configuration $\gamma=[111]$.
However, as shown by the \qvals, at an FDR of 0.05, the tissue-by-tissue analysis calls the eQTL only in fibroblasts.

Given the disagreement between the results from our novel framework and the original analyses of these data, we checked 
the plausibility of our results by applying a previously-used method for examining pairs of tissues to these data (\cite{Nica2011Architecture}).
This analysis takes the best eQTL in each gene identified in one tissue, and then estimates the proportion of these ($\pi_1$) that are also eQTLs in a second tissue 
(by applying Storey's method \cite{Storey2003Statistical} to their nominal \pvals in that second tissue, uncorrected for multiple comparisons). Unlike the tissue-by-tissue
analysis above, this approach avoids thresholding of \pvals, and makes some allowance for incomplete power. 
However, unlike our framework, this approach can only be applied to compare pairs of tissues. Applying this approach to each pair yielded a mean estimate of
$\pi_1 \approx 88\%$ (range $77\%$ to $94\%$), broadly consistent with our qualitative conclusion that a substantial proportion of eQTLs are shared among tissues.

\section*{Discussion}
In this work, we have presented a statistical framework for 
analyzing and identifying eQTLs, combining data from multiple tissues. Our approach
considers a range of alternative models, one for each possible configuration of eQTL sharing among
tissues. We compute Bayes Factors that quantify the support in the data for each possible configuration,
and these are used both to develop powerful test statistics for detecting genes that have an eQTL in at least one tissue
(by Bayesian model averaging across configurations), and to identify the tissue(s) in which these eQTLs are active
(by comparing the Bayes factors for different configurations against one another).
Our framework allows for heterogeneity of eQTL effects among tissues in which the eQTL is active, for different variances of gene expression measurements in each tissue,
and for intra-individual correlations that may exist due to samples being obtained from the same individuals.
For eQTL detection, our framework provides consistent, and sometimes substantial,
gains in power compared to a tissue-by-tissue analysis and ANOVA or simple linear regression.
Concerning the tissue specificity of eQTLs, our framework efficiently borrows information across genes to estimate configuration proportions, and then uses these estimates to assess the evidence for each possible configuration. When re-analyzing the gene expression levels in three cell types from 75 individuals (\cite{Dimas2009Common}), we found
that there appears to be a substantial amount of sharing of eQTLs among tissues, substantially more than suggested by the original analysis.

In the next few years, we expect that expression data will be available on large numbers of diverse tissue types in sufficient
sample sizes to allow eQTLs to be mapped effectively (for example, the NIH GTEx project aims to collect such data).
The methods presented here represent a substantive step towards improved analyses that 
fully exploit the richness of these kinds of data. However, we also see several directions for potential extensions
and improvements. First, our current framework can only partially deal with the challenges of large numbers of tissues.
Specifically, because with $S$ tissues, there are $2^S$ possible configurations of eQTL sharing among tissues, some of our
current methods, which consider all possible configurations, will become impractical for moderate $S$ (speculatively, above about 10, perhaps).
Our test statistic $\bmalite$ partially addresses this problem, by allowing for heterogeneity while averaging over only $S+1$
configurations, which is practical for very large $S$. 
Our simulation results suggest that $\bmalite$ is a powerful test statistic for identifying SNPs that are an eQTL in at least one tissue.
However our preferred approach for identifying {\it which} tissues such SNPs are active in involves a hierarchical model that
estimates the frequency of different patterns of sharing from the data, and this hierarchical model scales poorly with
$S$. In particular, having a separate parameter for each possible configuration is unattractive (both statistically and computationally) 
for large $S$, and alternative approaches will likely be required. There are several possible ways forward here: for example, one would
be to reduce the number of distinct configurations by clustering ``similar" configurations together; another would be to focus
less on the discrete configurations, and instead to focus on modeling heterogeneity in effect sizes in a continuous way - perhaps
using a mixtures of multivariate normal distributions with more complex covariance structures than we allow here.
We expect this to remain an area of active research in the coming years, especially since these
types of issues will likely arise in many genomics applications involving multiple cell types, and not only in eQTL mapping.

Another important issue to address is that most future expression data sets will likely be collected by RNA-seq, which
provides count data that are not normally distributed. Previous eQTL analyses 
of RNA-seq (e.g.\cite{Pickrell2010Understanding}) have nonetheless performed eQTL mapping using a normal model, by first transforming 
(normalized) count data at each gene to the quantiles of a standard normal distribution. Although this approach
would not be attractive in experiments with small sample sizes, with the moderate to large sample sizes typically
 used in eQTL mapping experiments this approach works well. As a first step, this approach could also
be used to apply our methods to count data. However, ultimately it would seem preferable to replace the normal model 
with a model that is better adapted to count-based data, perhaps a quasi-Poisson generalized linear model (\cite{Sun2012Statistical});
Bayes Factors under these models could be approximated using Laplace approximations, similar to the approximations used here
for the normal model \cite{Wen2011Bayesian}. The quasi-Poisson model has the advantage over the normal transformation
approach that it preserves the fact that there is more information about eQTL effects in tissues where a gene is high expressed
than in tissues where it is low expressed. This information is lost by normal transformation. In our analyses here
we addressed this by analyzing only genes that were robustly expressed in all tissues, but this is sub-optimal,
and will become increasingly unattractive as the number of tissues grows.

\section*{Methods}
\section*{Materials and Methods}

Software implementing our methods are available on the website \verb+http://stephenslab.uchicago.edu/software.html+.


\subsection*{Bayesian Methods for Mapping Multiple-tissue eQTLs}

\subsubsection*{Models for Multiple-tissue eQTLs}

For each tissue, we model the potential genetic association between a target SNP and the expression levels of a target gene by the simple linear regression model (\ref{simple.reg.model}). In vector form, this model is represented by
\begin{equation}
  \label{reg.model}
  \yv_s = \mu_s \lv + \beta_s \gv_s + \ev_s,~\ev_s \sim \mathcal{N}(0,\sigma_s^2 I),
\end{equation}  
where $s$ indexes one of the $S$ tissue types examined and the vectors $\yv_s, \gv_s$ and $\ev_s$ denote the expression levels, the genotypes of the samples and the residual errors respectively for the $s^{\text{th}}$ tissue type.
The intercept term, $\mu_s$, and the residual error variance, $\sigma_s^2$ are allowed to vary with tissue type.
The regression coefficient $\beta_s$ denotes the effect of the eQTL in tissue $s$, but we follow \cite{Wen2011Bayesian, Servin2007Imputationbased} in using the (unitless) standardized regression coefficient $b_s := \beta_s / \sigma_s$, 
as the main measure of effect size. As a result, inference is invariant to scale transformations of the response variables ($\yv_s$) within each tissue.

When the tissue samples are taken from the same individuals we allow that the observations on the same individual may be correlated with one another.
Specifically, let $E := (\ev_1\,\cdots\,\ev_s)$  denote the $N \times S$ matrix of residual errors, the we assume it to follow a matrix-variate normal (MN) distribution, i.e.,
\begin{equation}
  E \sim {\rm MN}(0, I, \Sigma).
\end{equation}    
That is, the vectors $(\epsilon_{1i},\dots,\epsilon_{Si})$ are independent and identically distributed as $\mathcal{N}(0, \Sigma)$.
The (unknown) $S \times S$ covariance matrix $\Sigma$ 
quantifies the correlations between the $S$ tissues; it can vary from gene to gene and is estimated from the data (see below). 
[When the tissue samples are collected from different individuals then we assume
their error terms are independent; methods
for this case are given in \cite{Wen2011Bayesian}.]

\subsubsection*{Prior on effect sizes}

A key component of our Bayesian model is the distribution $p(\bv | \gamma, \theta)$, where $\theta$ denotes hyper-parameters that are to be specified or estimated from the data. (In the main text we used $p(\beta | \gamma, \theta)$ to simplify exposition, but we actually work with the standardized effects $\bv$.) Of course, if $\gamma_s=0$ then
$b_s=0$ by definition. So it remains to specify the distribution of the 
remaining $b_s$ values for which $\gamma_s=1$.

We use the distribution from \cite{Wen2011Bayesian} (see also \cite{Lebrec2010Dealing,Han2011RandomEffects}), which provides a flexible way to model the heterogeneity of genetic effects of an eQTL in multiple tissues.
Specifically, \cite{Wen2011Bayesian} consider a distribution $p(\bv | \phi, \omega, \gamma)$, with two hyper-parameters, $\phi,\omega$, 
in which the non-zero effects are normally
distributed about some mean $\bar b$, which itself is normally distributed:
\begin{equation}
  \label{het.prior}
  b_s  | \bar b, \gamma_s=1 \sim \mathcal{N}(\bar b, \phi^2),
\end{equation}    
and 
\begin{equation}
  \label{avg.prior}
  \bar b \sim \mathcal{N}(0, \omega^2).
\end{equation}
Note that $\phi^2+\omega^2$ controls the variance (and hence the
expected absolute size) of $b_s$, and $\phi^2/(\phi^2+\omega^2)$ controls
the heterogeneity (indeed, $\omega^2/(\phi^2+\omega^2)$ is the correlation of $b_s, b_{s'}$ for different subgroups $s \neq s'$). 
If $\phi^2=0$ then this model corresponds to the ``fixed effects" model in which the effects in all subgroups are equal (e.g. \cite{Han2012Interpreting}).

To allow for different levels of effect size and heterogeneity, \cite{Wen2011Bayesian} use a fixed grid of values $\{(\phi_i, \omega_i): i = 1,\dots,L\}$, with the $i$th grid point having weight $w_i$. Thus
\begin{equation}
p(\bv | \gamma, \theta)= \sum_i w_i p(\bv | \phi_i,\omega_i,\gamma).
\end{equation}
In all our applications here we consider the grid of values fixed, and 
treat the weights $w_1,\dots,w_L$ as hyper-parameters (so $\theta= (w_1,\dots,w_L)$), which can be either fixed or estimated.

\subsubsection*{Choice of grid for $(\phi,\omega)$}

We define a grid of values for $(\phi,\omega)$ by specifying a set $E$ of values for the average effect size, $\omega^2+\phi^2$, and a set $H$ of values for the heterogeneity $\phi^2/(\phi^2+\omega^2)$, and then taking the grid to be all $L = |E| \times |H|$ possible combinations of values. For all methods we use $E= \{0.1^2, 0.2^2, 0.4^2, 0.8^2, 1.6^2\}$, which is designed to span a wide range of eQTL effect sizes. For $\bma$ and $\bmahm$ we allow for only a limited range
of heterogeneity: $H= \{0,0.25\}$. In this way we assume that when the eQTL is
present in multiple tissues, it has a similar (but not necessarily identical)
effect in each tissue. For $\bmalite$ we allow a much wider range of heterogeneity: $H=\{0,0.25,0.5,0.75,1\}$. The rationale here is that the large heterogeneity values will help capture eQTLs that are present in only a subset of tissues, a feature that is not otherwise captured by $\bmalite$ as
it averages over a small number of configurations. 

\subsubsection*{Choice of weights $\wv$ and $\etv$}

Let $|\gamma|:= \sum_s \gamma_s$ denote the number of elements of $\gamma$ that are equal to 1 (i.e.~the number of tissues in which the eQTL is
active in configuration $\gamma$). 

For $\bma$ we fix the weights $\etv$ so that they put weight $1/S$ on all $S$ possible values for $|\gamma|$, and, conditional on $|\gamma|$, put equal weight on all $S\choose|\gamma|$ configurations with that value for $|\gamma|$. Thus $\eta_\gamma = (1/S) {S\choose|\gamma|}^{-1}$. In addition we fix the grid weights $\wv$ to be equal on all grid values.

For $\bmalite$ we put non-zero weights $\etv$ on
only the consistent configuration ($|\gamma|=S$) 
and configurations with an eQTL in a single tissue ($|\gamma| = 1$). We set $\etv$ so that it puts weight 0.5 on each of $|\gamma|=1$ and $|\gamma|=S$.
Conditional on $|\gamma|=1$ we assume all $S$ possibilities are equally likely.
Thus $\eta_\gamma = 0.5$ if $\gamma=[111\dots1]$ and 0.5/S if $|\gamma|=1$. Again, we fix the grid weights $\wv$ to be equal on all grid values (but with
the larger grid for heterogeneity described above).

 For $\bmahm$ we estimate the weights $\wv, \etv$ from the data using
a hierarchical model to combine information across genes, as described below.

\subsubsection*{Bayes Factor Computation}

To complete model specification, we use (limiting, diffuse) prior distributions
for the nuisance parameters $\mu_s$ and $\Sigma$, as in \cite{Wen2012Bayesian}. Under these priors we can compute the Bayes Factor
$\BF_\gamma$ in (\ref{eqn:bfgamma}) using 
\begin{equation} 
   \label{bf.avg}
  \BF_\gamma = \sum_{j=1}^M w_j \BF_\gamma(\phi_j, \omega_j)
\end{equation}
where $\BF_\gamma(\phi_j, \omega_j)$ is given by
\begin{equation} \label{eqn:bfphi}
\BF_\gamma(\phi, \omega) = \frac{p(Y | G, \phi, \omega,\gamma)}{p(Y | G, H_0)} = \frac{\int p(Y \,|\,G, \mu,b,\Sigma)p(\mu,\Sigma) p(b|\gamma, \phi,\omega) \,db \, d\mu \, d\Sigma}{ \int p(Y \,|\,G, \mu,b=0,\Sigma)p(\mu,\Sigma) \, d\mu \, d\Sigma}
\end{equation}    
where $Y$ and $G$ denote the collection of expression levels and genotypes for a target gene-SNP pair across all tissue types respectively.
We use analytic approximations for these Bayes Factors based on Laplace approximation, given in \cite{Wen2011Bayesian, Wen2012Bayesian}. In particular, we use the approximation  which in functional forms is connected to Frequentist's score statistic.

\subsection*{Bayesian Hierarchical Model}

For $\bmahm$ we use a hierarchical model, similar to \cite{Veyrieras2008Highresolution,Maranville2011Interactions}, which combines information across genes, to estimate the grid weights $\wv$'s and configuration weights $\etv$'s. Following both 
\cite{Veyrieras2008Highresolution,Maranville2011Interactions} we make the simplifying assumption that
each gene has at most one eQTL (which may be active in multiple tissues),
and that each SNP is equally likely to be the eQTL. Let $m_k$ be the number of SNPs in the {\it cis}-region for gene $k$. Then, if $\BF^{k,v}_\gamma(\phi,\omega)$ denotes the Bayes Factor (\ref{eqn:bfphi}) computed for SNP $v$ in gene $k$, the ``overall Bayes Factor" measuring the evidence
for an eQTL in gene $k$, $\BF^k$, is obtained by averaging over the possible eQTL SNPs, the possible configurations $\gamma$, and the grid of values for $\phi,\omega$, weighting by their probabilities:
\begin{equation}
\BF^{k}(\etv,\wv) = \frac{p(\text{data at gene $k$} | \text{gene contains eQTL})}{p(\text{data at gene $k$}  | \text{gene contains no eQTL})} =
(1/m_k) \sum_{v=1}^{m_k} \sum_i \sum_\gamma \eta_\gamma w_i \BF^{k,v}_\gamma(\phi_i,\omega_i).  
\end{equation}
Furthermore, if we let $\pi_0$ denote the probability that each gene 
follows the null (i.e.~contains no eQTL) 
then the likelihood for gene $k$, as a function of $\pi_0, \etv,\wv$, is given by
\begin{align}
L_k(\pi_0, \etv,\wv) & = (1-\pi_0)p(\text{data at gene $k$} | \text{gene contains eQTL}) + \pi_0 p(\text{data at gene $k$} | \text{gene contains no eQTL}) \\
&\propto (1-\pi_0) \BF^k + \pi_0
\end{align}
The overall likelihood for our hierarchical model is obtained by multiplying
these likelihoods across genes:
\begin{equation}
L(\pi_0, \etv, \wv) = \prod_k L_k (\pi_0,\etv,\wv).
\end{equation}
Note that although the expression levels for different genes
are not independent, because the SNPs being tested in different genes are
mostly independent this independence assumption for the likelihoods across genes
is a reasonable starting point.
We have developed both an EM algorithm to estimate the parameters
$(\pi_0, \etv, \wv)$ by maximum likelihood
(see Supplementary information).

\subsubsection*{Relaxation of ``one {\it cis}-eQTL per gene'' Assumption}

To relax the ``one {\it cis}-eQTL per gene'' assumption we adopt the following
procedure. First we compute the posterior probability of each SNP being the sole eQTL for each gene (i.e. only allowing one {\it cis}-eQTL per gene) with a set of default parameters, and use these to identify the top SNP for each gene (i.e.~the one with the largest posterior probability of being the eQTL).

For each gene, separately in each tissue, we compute the residuals of its expression level after regressing out the effect of the top SNP.
If these residuals are strongly associated with a SNP then this is evidence
for that SNP being a second independent eQTL for that gene. Therefore,
to allow for the potential for more than one eQTL per gene we treat these residuals as defining a second set of ``artificial" expression data for each gene and each tissue, and  fit the hierarchical model using both the original and the artificial expression data.


\subsection*{Simulation procedures}

For our simulations, when simulating SNP-gene pairs, the genotypes at each SNP in each individual were simulated as Binomial(2,0.3): that is, with minor allele frequency 30\% and assuming Hardy-Weinberg equilibrium. Phenotypes with eQTLs were simulated, with effect size based on an expected proportion of variance explained (PVE) of 20\%;  (see supplementary text S1). 
For Figures \ref{fig:fig1}A and \ref{fig:fig1}B)
the error variances (one per tissue) were all equal to 1.
For Figure \ref{fig:fig1}C the error variances were randomly drawn from $\{1, 1.5 ,2 \}$, all equally likely.

\subsection*{The ANOVA/LR method}

The ANOVA/LR method uses the same linear model
as our Bayesian methods (\ref{simple.reg.model}), except that the residual errors $\sigma_s$ are assumed to be equal across tissues $s$.
Within this model we tested the global null hypothesis
($\beta_s=0$ for all $s$) using an $F$ test comparing the null
model with the unconstrained alternative ($\beta_s$ unconstrained).


\subsection*{Preprocessing of the data set from Dimas \textit{et al.}}

The phenotypes from Dimas \textit{et al.} (\cite{Dimas2009Common}) were retrieved from the Gene Expression Omnibus (GSE17080).
We mapped the 22,651 non-redundant probes to the hg19 human genome reference sequence (only the autosomes) using \verb+BWA+ (\cite{Li2009Fast}), kept 19,965 probes mapping uniquely with at most one mismatch, and removed the probes overlapping several genes from Ensembl.
This gave us 12,046 genes overlapped by 16,453 probes.
For genes overlapped by multiple probes, we chose a single probe at random.
In our analyses we considered only genes that were 
robustly expressed in all tissues.
A gene was considered robustly expressed in a given tissue if its mean expression level across individuals in this tissue was larger than or equal to the median expression level of all genes across all individuals in this tissue.
As a result, we focused on 5012 genes.

Genotypes were obtained from the European Genome-phenome Archive
(EGAD00000000027).
We extracted the genotypes corresponding to the 85 individuals for which we had phenotypes and converted the SNP coordinates to the hg19 reference using \verb+liftOver+ (\cite{Hinrichs2006UCSC}).
To detect outliers, we performed a PCA of these genotypes using individuals from the CEU, CHB, JPT and YRI populations of the HapMap project using \verb+EIGENSOFT+ (\cite{Price2006Principal}).
As in the original study, we identified 10 outliers and removed them from all further analyses, which were therefore performed on 75 individuals.

Gene expression measurements suffer 
from various confounders, many of which may be unmeasured (\cite{Leek2010Tackling}), but which can be corrected for using methods such as principal components analysis (PCA).
Following \cite{Pickrell2010Understanding}, we applied PCA in each tissue separately on the 5012 $\times$ 75 matrix of expression levels of each gene in each individual.
We sorted principal components (PCs) according to the proportion of variation in the original matrix they explain, and selected PCs so that adding another PC would explain less than 0.0025\% of the variation.
As a result, this procedure identified 16 PCs in Fibroblasts, 7 in LCLs and 15 in T-cells.
We then regressed out these PCs from the original matrix of gene expression levels, and used the residuals as phenotypes for all analyses.

All methods we compared assume that the errors are distributed according to a Normal distribution. Before analysis we therefore
rank-transformed the expression levels at each gene to the quantiles
of a standard Normal distribution (\cite{Servin2007Imputationbased}).


\subsection*{Permutation procedures}

On the data set from Dimas \textit{et al.}, we assessed the performance of two methods, the tissue-by-tissue analysis and the BMA joint analysis, by comparing the number of genes identified as having at least one eQTL in any tissue, at a given FDR. For each method, we defined a test statistic, which was computed for each gene.
For the tissue-by-tissue analysis, the test statistic is the minimum \pval of the linear regressions between the given gene and each \textit{cis} SNP in each tissue (so the minimum is taken across all SNPs and all tissues). For the BMA joint analysis, the test statistic is the average of the Bayes Factors for the given gene and each \textit{cis} SNP. (When applying the tissue-by-tissue analysis
to test for eQTLs in a single tissue, the test statistic is the minimum \pval of the linear regressions between the given gene and each \textit{cis} SNP in that tissue.)

In each case we converted the test statistic to 
a \pval for each gene, testing the null hypothesis that
the gene contains no eQTL in any tissue, by comparing the observed
test statistic with the value of the test statistic obtained on permuted
data obtained by permuting the individuals labels (using the
same permutations in each tissue to 
preserve any intra-individual correlations between gene expression in different tissues). Specifically, let $P$ denote the total number of permutations (we used $P=10^4$), $T_g$ the value of the test statistic for gene $g$ on the non-permuted data, and $T_g^{(i)}$ the value of the test statistic on the $i^{th}$-permuted data.
The \pval for gene $g$ from the tissue-by-tissue analysis is: $(1 + \sum_{i=1}^P \mathbf{1}_{T_g^{(i)} \le T_{g}}) / (1 + P)$.
For the BMA joint analysis, the \pval is: $(1 + \sum_{i=1}^P \mathbf{1}_{T_g^{(i)} \ge T_{g}}) / (1 + P)$.
Note that permutations were performed for each gene, since the null distribution of the test statistic will vary across genes (not least
because the genes have different numbers of SNPs in their \textit{cis} candidate region; see supplementary figure \ref{fig:fig_s1}).

From the \pval calculated for each gene we estimate $q$ values \cite{Storey2003Statistical}
using the \verb+qvalue+ package, and determine the 
number of genes having at least one eQTL in any tissue at an FDR of $\alpha$ by computing the number of genes with $q \le \alpha$.

When performing the tissue-by-tissue analysis on a single tissue, we performed the permutations in each tissue separately.

\section*{Acknowledgments}
We thank John Zekos for technical support 
and members of the Przeworski, Pritchard and Stephens labs for helpful discussions.

\bibliography{paper-eQtlBma_main}

\appendix

\section{Computational algorithm for fitting hierarchical model}

For the hierarchical model described in the main text, our primary interest is making inference on the parameter set $\Theta = (\pi_0, \etv, \lav)$. Here, we give details of an algorithm for inferring $\Theta$, via maximum likelihood estimation based on the EM algorithm.

\subsection{Notations}

Throughout this section, we adopt the following additional notations. For gene $k$, we use a latent binary indicator $z_k$ to denote if there is any eQTL in its {\it cis}-region for any tissue type, in particular,
\begin{equation}
  \Pr(z_k = 1) = 1 - \pi_0;
\end{equation}
We use a latent random indicator $m_k$-vector $\sv_k$ to denote the true eQTL SNP conditional on $z_k=1$ and let $s_{kp}$ denote the $p$-th entry of $\sv_k$. The ``one {\it cis} eQTL per gene'' assumption restricts $\sv_k$ can have at most one entry equaling 1 (with the remaining entries being 0). By this definition,  
\begin{equation}
  \Pr( \sv_k = {\bf 0}  | z_k = 0)  = 1,
\end{equation}
and we also make the simplifying assumption that
\begin{equation}
  \Pr( s_{kp} = 1  | z_k = 1)  =  \frac{1}{m_k}.
\end{equation}
Furthermore, for gene $k$ and SNP $p$, we index all configurations and use a $(2^S-1)$-dimension latent indicator vector $\cv_{kp}$ to denote the actual configuration for the gene--SNP pair. In case the SNP is not an eQTL,   
 \begin{equation}
  \Pr( \cv_{kp} = {\bf 0}  | s_{kp} = 0)  = 1.
\end{equation}
Otherwise, we assume the $j$th configuration is active with prior probability
  \begin{equation}
  \Pr( c_{kpj} = 1  | s_{kp} = 1)  = \eta_j.
\end{equation}
Joining the column vectors $\cv_{kp}$ for all $m_k$ SNPs, we obtain a latent $(2^S-1) \times m_k$ random matrix $C_k$.  
Finally, we use the latent $L$-vector $\wv_{kp}$ indicate the actual prior effect size for active tissue types for the pair of gene $k$ and SNP $p$. The $m$-th entry of the indicator is denoted by $w_{kpm}$, for which we assume
prior probability
\begin{equation}
  \Pr(\wv_{kp}= {\bf 0} | s_{kp} =0 ) =  1,
\end{equation}
and
\begin{equation}
	\Pr(w_{kpm}=1 | s_{kp} =1 ) =  \lambda_m. 
\end{equation}
Joining the column vectors $\wv_{kp}$ for all $m_k$ SNPs, we obtain a latent $L \times m_k$ random matrix $W_k$

\subsection{Maximum Likelihood Inference based on EM algorithm}

In the maximum likelihood framework, we treat latent variables $z_k, \sv_k, \cv_k$ and $\wv_k, k=1,\dots,G$ as missing data and apply the EM algorithm. 

For a total number of $G$ genes, let $\zv = (z_1,\dots,z_G), \Sv = (\sv_1,\dots, \sv_G), \Cv = (C_1,\dots,C_G) $ and $\Wv = (W_1,\dots W_G) $ denote the complete collection of latent variables. Let $\Yv = (\Yv_1, \dots \Yv_G)$ and $\Gv = (G_1, \dots, G_G)$ denote the complete set of observed data. Based on the hierarchical model described in previous section, we can write out the complete data log-likelihood as follows,    
\begin{equation}
 \label{cll}  
  \begin{aligned}
    \log ~&p(\Yv,\zv,\Sv,\Cv,\Wv|\Gv,\Theta) = \\
         &\sum_k (1-z_k)\log \pi_0 + \sum_k z_k \log (1-\pi_0) \\
         & + \sum_{k,p}z_ks_{kp}\log \frac{1}{m_k} + \sum_{k,p,j}z_k s_{kp} c_{kpj}\log \eta_j+ \sum_{k,p,m}z_k s_{kp} w_{kpm} \log \lambda_m  \\
        & + \sum_{k,p,j,m} z_k s_{kp} c_{kpj} w_{kpm} \cdot {\rm BF}_{kpjm} + \sum_k \log p^0_k.
    \end{aligned}
  \end{equation}
In (\ref{cll}), $p^0_k$ denotes the likelihood of the null model for gene $k$, i.e.,  
\begin{equation}
\label{p_null}
  p^0_k := p(\Yv_k | z_k=0)
\end{equation}
and  
 \begin{equation}
 \label{bf_kpjm}
 {\rm BF}_{kpjm} =  \frac{P(\Yv_k | z_k=1,s_{kp}=1, c_{kpj}=1, w_{kpm}=1,\Gv_k,\Theta)}{p^0_k}
\end{equation}
is the Bayes Factor (pre-)computed for a fully specified alternative model.

The EM algorithm searches for maximum likelihood estimate of $\Theta$, by iteratively performing an expectation (E) step and a maximization (M) step. 

In the E-step, for the $t$-th iteration, we evaluate the expectation of complete data log-likelihood (\ref{cll}) conditional on current estimate of parameter $\Theta^{(t)}$, $\Gv$ and $\Yv$. The computation is straightforward, for example,
\begin{equation}
\begin{aligned}
    {\rm E}(z_k | \Yv_k, \Gv_k, \Theta^{(t)})  &= \Pr(z_k=1 | \Yv_k, \Gv_k,\Theta^{(t)}) \\ 
    & = \frac{\Pr(z_k=1|\Theta^{(t)})\cdot p(\Yv_k|z_k=1,\Gv_k,\Theta^{(t)})}{p(\Yv_k|\Gv_k,\Theta^{(t)})} \\
    & =  \frac{(1-\pi_0^{(t)}){\rm BF}^{(t)}_{k}}{\pi_0^{(t)} + (1-\pi_0^{(t)}){\rm BF}^{(t)}_{k}},
  \end{aligned}
\end{equation}
similarly,
\begin{align}
  & {\rm E}(z_k s_{kp} | \Yv_k, \Gv_k, \Theta^{(t)})  = \frac{(1-\pi_0^{(t)})\frac{1}{m_k}{\rm BF}^{(t)}_{kp}}{\pi_0^{(t)} + (1-\pi_0^{(t)}){\rm BF}^{(t)}_{k}},\\
  & {\rm E}(z_k s_{kp} c_{kpj} w_{kpm}  | \Yv, \Gv, \Theta^{(t)}) = \frac{(1-\pi_0^{(t)})\frac{1}{m_k}\eta_j^{(t)}\lambda_m^{(t)}{\rm BF}_{kpjm}}{\pi_0^{(t)} + (1-\pi_0^{(t)}){\rm BF}^{(t)}_{k}},
\end{align}

where 
\begin{equation}
 \label{bf_k}
  \begin{aligned}
   {\rm BF}^{(t)}_{k} &=  \frac{p(\Yv_k|z_k=1,\Gv_k,\Theta^{(t)})}{p^0_k} \\
               &=\sum_{p,j,m} \frac{1}{m_k} \eta_j^{(t)} \lambda_m^{(t)} {\rm BF}_{kpjm},
  \end{aligned}
\end{equation}

and 

\begin{equation}
  \label{bf_kp}
  \begin{aligned}
  {\rm BF}^{(t)}_{kp} &= \frac{p(\Yv_k|z_k=1,s_{kp}=1,\Gv_k,\Theta)}{p^0_k} \\
               &=\sum_{j,m} \eta_j^{(t)} \lambda_m^{(t)} {\rm BF}_{kpjm},
  \end{aligned}
\end{equation}

In the M-step, we find a new set of estimates, $\Theta^{(n+1)}$, by maximizing the conditional expectation ${\rm E}\left(\log p(\Yv,\zv,\Sv,\Cv,\Wv|\Gv,\Theta) | \Yv,\Gv,\Theta^{(t)}\right)$. In this case, the simultaneous maximization can be performed analytically. In particular, 
\begin{align}
  &\pi_0^{(t+1)} = \frac{1}{g}\sum_{k=1}^g \frac{\pi_0^{(t)}}{\pi_0^{(t)}+(1-\pi_0^{(t)}){\rm BF}_k^{(t)}},\\
  & \eta_j^{(t+1)} = \frac{ \sum_{k,p,m}\frac{\gamma_{kp}^{(t)}\lambda_m^{(t)}{\rm BF}_{kpjm}}{\pi_0^{(t)} + (1-\pi_0^{(t)}){\rm BF}^{(t)}_{k}}\cdot \eta_j^{(t)} }{\sum_{j'} \bigg(\sum_{k,p,m}\frac{\gamma_{kp}^{(t)}\lambda_m^{(t)}{\rm BF}_{kpj'm}}{\pi_0^{(t)} + (1-\pi_0^{(t)}){\rm BF}^{(t)}_{k}}\cdot  \eta_{j'}^{(t)} \bigg)} ,\\
\intertext{and}
  &\lambda_m^{(t+1)} = \frac{ \sum_{k,p,j}\frac{\gamma_{kp}^{(t)}\eta_j^{(t)}{\rm BF}_{kpjm}}{\pi_0^{(t)} + (1-\pi_0^{(t)}){\rm BF}^{(t)}_{k}}\cdot\lambda_m^{(t)} }{\sum_{m'} \bigg(\sum_{k,p,j}\frac{\gamma_{kp}^{(t)}\eta_j^{(t)}{\rm BF}_{kpjm'}}{\pi_0^{(t)} + (1-\pi_0^{(t)}){\rm BF}^{(t)}_{k}} \cdot\lambda_{m'}^{(t)}    \bigg)}.
\end{align}

Typically, we initiate the EM algorithm by setting $\Theta^{(0)}$ to some random values and running iterations until some pre-defined convergence threshold is met (In practice, we monitor the increase of the  the log-likelihood function between successive iterations, and stop the iterations as the increment becomes sufficiently small.).

We construct profile likelihood confidence intervals for estimated parameters. For example, a $(1-\alpha)\%$ profile likelihood confidence set for $\pi_0$ is built as 
\begin{equation}
  \{\pi_0: \log p(\Yv | \pi_0,\hat \etv, \hat \lav, \Gv) > \log p(\Yv | \hat \pi_0,\hat \etv, \hat \lav, \Gv) - \frac{1}{2}Z^2_{(1-\alpha)} \},
\end{equation}
where $\hat \pi_0, \hat \etv, \hat \lav$ are MLEs obtained from the EM algorithm.


\section{Supplements for the simulations}

\subsection{Simulate eQTL data via the proportion of variance explained}

For a given gene-SNP pair at a time, we simulate data in $S$ tissues according to a particular configuration.
In a given tissue $s \in \{1,\ldots,S\}$ for which the SNP is an eQTL ($\beta_s \neq 0$), let's define the proportion of variance in phenotype explained by the genotype:
\[
PVE_s (\beta_s, \sigma_s) = \frac{V(X \beta_s)}{V(X\beta_s) + \sigma_s^2}
\]
When working with standardized effect sizes $b_s = \beta_s / \sigma_s$:
\[
PVE_s = \frac{V(Xb_s)}{V(Xb_s) + 1}
\]
As stated elsewhere (\cite{Guan2011Bayesian}), we approximate the expectation of the PVE via a ratio of expectations, noted $h$:
\[
h = \frac{E[V(Xb_s)]}{E[V(Xb_s)] + 1}
\]

We assume that the genotypes are drawn from a Binomial distribution with parameters 2 and $f$, the minor allele frequency, so that $E[V(X)] = 2 f (1-f)$.
Moreover, as we assume $b_s | \bar{b} \sim N(\bar{b}, \phi^2)$ and $\bar{b} \sim N(0, \omega^2)$, the marginal effect size is $b_s \sim N(0,\phi^2 + \omega^2)$.
We can hence approximate $b_s^2$ by its variance.
Therefore:
\[
h = \frac{(\phi^2 + \omega^2) \times 2 f (1-f)}{(\phi^2 + \omega^2) \times 2 f (1-f) + 1}
\]

By fixing $h$ (eg. 20\%) as well as the minor allele frequency (eg. 30\%), we obtain:
\[
\phi^2 + \omega^2 = \frac{h}{(1 - h) \times 2 f (1-f)}
\]
Now if we fix the heterogeneity in effect sizes (eg. $\phi^2 / (\phi^2 + \omega^2) = 20\%$), we can deduce $\phi^2$ and then $\omega^2$.
We can hence draw $\bar{b}$ and then each $b_s|\bar{b}$.

Once we have them, it is straightforward to simulate the phenotype of the $i^{th}$ individual in the $s^{th}$ tissue:
\[
y_{is} = b_s \sigma_s g_{i} + \mathcal{N}(0,\sigma_s^2)
\]
with $\sigma_s$ being fixed at 1 for instance.


\subsection{Implement the ANOVA/LR model in R}

For each gene-SNP pair, the expression levels from all $N$ individuals in all $S$ tissues are recorded into a vector \verb+y+ of length $N \times S$.
The genotypes are appropriately repeated $S$ times into a vector \verb+xg+, and the tissue indicators are appropriately recorded into a vector \verb+xs+.
We can then use the ANOVA/LR model to test if there is an effect of the genotype with the following commands:

\begin{verbatim}
m1 <- lm(y ~ xs)
m2 <- lm(y ~ xs * xg)
pval <- anova(m1, m2)[[6]][2]
\end{verbatim}


\subsection{Calculate the empirical FDR from simulated eQTL data}

For a simulated data set of $G$ gene-SNP pairs, let $z_g$ be the test statistic of a given pair with $g = 1, \ldots, G$.
For the tissue-by-tissue method, we take as test statistic the minimum P-value across tissues.
For the Bayesian method, the test statistic is the Bayes Factor.
For the ANCOVA, the test statistic is the P-value of the genotype effect with interaction.

All gene-SNP pairs can be classified as in the following table (\cite{Storey2003Statistical}):
\begin{center}
\begin{tabular}{llll}
            &  Called eQTL  &  Not called  &  Total  \\
 True null  &  F            &  G$_0$ - F   &  G$_0$  \\
 True eQTL  &  T            &  G$_1$ - T   &  G$_1$  \\
 Total      &  S            &  G - S       &  G      \\
\end{tabular}
\end{center}

As we simulate data, we know which pairs are true eQTLs.
By fixing the empirical false discovery rate ($FDR_e$) at 5\%, we can find the corresponding cutoff $c$ on the test statistics, and from there calculate the true positive rate (TPR) at this cutoff:

$TPR(c) = T(c) / G_1$ with $c$ such that $FDR_e(c) = F(c) / S(c) = 0.05$.

The following algorithm describes how to iteratively find the cutoff $c$ corresponding to the 5\% empirical FDR:

\vspace{2em}
\IncMargin{2em}
\begin{algorithm}[H]
\SetAlgoVlined
\DontPrintSemicolon
\KwData{test statistics $z_1,\ldots,z_G$}
\BlankLine
\If{P-values}{sort in increasing order: $z_{(1)} \le \ldots \le z_{(G)}$}
\ElseIf{Bayes Factors}{sort in decreasing order: $z_{(1)} \ge \ldots \ge z_{(G)}$}
\BlankLine
\ForEach{gene-SNP pair $g \leftarrow 1$ \KwTo $G$}{
$c\leftarrow z_{(g)}$\;
$s\leftarrow $number of called eQTLs at this cutoff $c$\;
$f\leftarrow $number of false positives among them\;
$fdr\leftarrow f / s$\;
\If{$fdr \ge 5\%$}{
$t\leftarrow $number of true positives among the called eQTLs\;
$tpr\leftarrow t / s$\;
exit\;
}
}
\end{algorithm}
\vspace{2em}

Between different methods, the empirical FDRs will always be 5\% (or slightly higher) but the TPRs and FPRs will be different, which allows us to compare the performance of the methods.


\section{Supplements for the analysis of the Dimas \textit{et al.} data set}

\subsection{Hierarchical model fed with Bayes Factors from residuals}

First we computed the Bayes Factors for each combination of grid values and configurations, one gene-SNP pair at a time.
Second, for each gene, we regressed out the effect of its best SNP, and we recomputed the Bayes Factors for the remaining SNPs using the residuals as phenotypes.
Third, we launched the hierarchical model with only the Bayes Factors obtained from the residuals.

If the ``at most one eQTL per gene'' assumption is reasonable for this data set, we would expect the estimated $\pi_0$ to be very high (meaning that the vast majority of genes have no eQTL), the lowest grid value to have the highest probability (meaning that the effect sizes are very small), and the credible intervals for the configurations to be very large (corresponding to high uncertainty).

This is indeed what we observe:

$\pi_0$: 0.963 [0.946, 1.000]

\begin{table}[!ht]
\centering
\begin{tabular}{|c|c|c|}
Grid value ($\phi^2, \omega^2$) & Posterior mean & 95\% credible interval \\
\hline
(0.01, 0.01) & 0.930 & [0.543, 1.000] \\
(0.01, 0.04) & 0.070 & [0.000, 0.459] \\
(0.01, 0.16) & 0.000 & [0.000, 0.107] \\
(0.01, 0.64) & 0.000 & [0.000, 0.036] \\
(0.01, 2.56) & 0.000 & [0.000, 0.019] \\
\end{tabular}
\end{table}

\begin{table}[!ht]
\centering
\begin{tabular}{|c|c|c|}
Configuration & Posterior mean & 95\% credible interval \\
\hline
100 & 0.316 & [0.000, 1.000] \\
010 & 0.330 & [0.000, 1.000] \\
001 & 0.027 & [0.000, 0.535] \\
110 & 0.250 & [0.000, 1.000] \\
101 & 0.020 & [0.000, 0.442] \\
011 & 0.029 & [0.000, 0.535] \\
111 & 0.028 & [0.000, 0.400] \\
\end{tabular}
\end{table}

\newpage

\subsection{Configuration proportions from all genes without removing expression PCs}

Similarly to what was done in the first analysis of this data set (\cite{Dimas2009Common}), we also analyzed the data set comprising all 12,046 genes, i.e.~without pre-selecting genes robustly expressed in all three tissues, and without removing expression PCs.
Here are the configuration proportions estimated by the EM algorithm:

\begin{table}[!ht]
\centering
\begin{tabular}{|c|c|}
Configuration & Hierarchical model \\
\hline
F-L-T & 0.793 [0.722, 0.878] \\
L-T & 0.071 [0.030, 0.129] \\
F-L & 0.000 [0.000, 0.015] \\
F-T & 0.000 [0.000, 0.015] \\
F & 0.052 [0.000, 0.109] \\
L & 0.058 [0.016, 0.115] \\
T & 0.025 [0.000, 0.068] \\
\end{tabular}
\end{table}

They are thus qualitatively similar to those obtained on the subset of genes robustly expressed in all three tissues and after having removed PCs (table 1 of the main text).

\setcounter{figure}{0}
\makeatletter 
\renewcommand{\thefigure}{S\@arabic\c@figure}

\begin{figure}[!ht]
\begin{center}
\includegraphics[width=\textwidth]{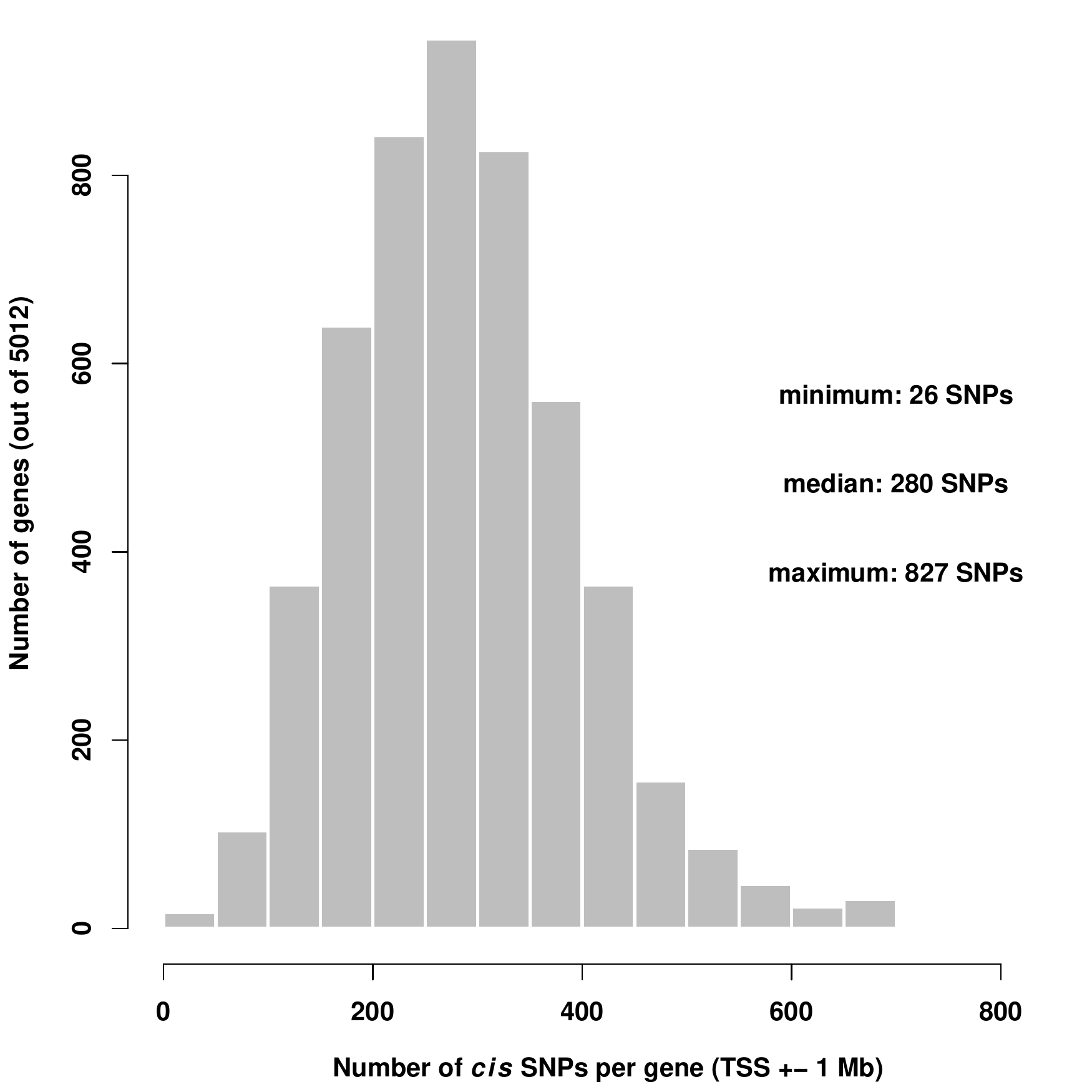}
\end{center}
\caption{
{\bf Distribution of the number of SNPs in the cis region of each gene.}
}
\label{fig:fig_s1}
\end{figure}


\end{document}